\documentclass[a4paper]{article}
\usepackage{graphicx}
\usepackage{multirow}
\usepackage{amsmath}
\usepackage{subfigure}
\usepackage{INTERSPEECH2022}
\usepackage{soul}
\usepackage{color, xcolor}

\title{LCSM: A Lightweight Complex Spectral Mapping Framework for Stereophonic Acoustic Echo Cancellation}
\name{Chenggang Zhang, Jinjiang Liu, Xueliang Zhang}
\address{
  Department of Computer Science, Inner Mongolia University, China}
\email{}

\begin{document}

\maketitle
\begin{abstract}
The traditional adaptive algorithms will face the non-uniqueness problem when dealing with stereophonic acoustic echo cancellation (SAEC). 
In this paper, we first propose an efficient multi-input and multi-output (MIMO) scheme based on deep learning to filter out echoes from all microphone signals at once. Then, we employ a lightweight complex spectral mapping framework (LCSM) for end-to-end SAEC without decorrelation preprocessing to the loudspeaker signals. Inplace convolution and channel-wise spatial modeling are utilized to ensure the near-end signal information is preserved. Finally, a cross-domain loss function is designed for better generalization capability. Experiments are evaluated on a variety of untrained conditions and results demonstrate that the LCSM significantly outperforms previous methods. Moreover, the proposed causal framework only has 0.55 million parameters, much less than the similar deep learning-based methods, which is important for the resource-limited devices.
\end{abstract}


\section{Introduction}
Stereophonic teleconferencing systems utilize spatial information to improve sound quality and realism compared with the single channel setting (i.e., one microphone and one loudspeaker). Traditional stereophonic acoustic echo cancellation (SAEC) methods usually remove the undesired echoes by estimating the acoustic echo paths between the stereophonic loudspeakers and microphones \cite{sondhi1993acoustic,sondhi1995stereophonic}. However, the non-uniqueness problem is caused by the strong linear correlation between the echo signals that are played out via loudspeakers. Furthermore, the echoes are not only auto-correlated but also cross-correlated, making the SAEC ill-posed and unstable \cite{BenestySAEC}.

In the past decades, a number of interchannel decorrelation strategies have been proposed to mitigate the problem, e.g., addition of uncorrelated noise \cite{gansler1998influence}, nonlinear preprocessing \cite{benesty1998stereophonic}, time-varying decorrelation \cite{ali1998stereophonic}, and non-intrusive method \cite{thune2013improved} etc. Lately, stereophonic acoustic echo suppression (SAES) without the decorrelation procedure was proposed in \cite{yang2012stereophonic, lee_stereophonic_2014}, which estimates the echo spectra from the stereo signals using the Wiener filter in the short-time Fourier transform (STFT) domain. However, the above methods are based on unrealistic prior assumptions of the signal model, which inevitably degrade sound quality and stereophonic spatial perception in practice.

More recently, deep learning-based methods which intrinsically avoid the non-uniqueness problem have been proposed for solving SAEC, and achieve impressive performance. Cheng \textit{et al.} \cite{cheng2021deep} employs the convolutional recurrent network (CRN) which takes the magnitude spectra of both far-end and microphone signals as inputs to estimate a mask for the near-end magnitude spectrum.  
In contrast to the mask estimation, Zhang and Wang \cite{zhang2021deep} proposed a method that learn the complex spectral mapping to directly estimate real and imaginary spectra of near-end speech, by using the complex spectrums of far-end and microphone signals as the CRN inputs. 
However, the typical CRN utilizes the convolution with stride operation, usually greater than one, possibly resulting in the spatial information confusion for the multi-channel hidden features representation and the independence property of each individual frequency bin is destroyed.

Recently, we proposed an inplace convolution recurrent neural network for single-channel AEC \cite{zhang2022complex}. In this paper, we address the SAEC problem. 
Specifically, we take all the far-end and microphone signals as one network input, and output the near-end signal in each microphone at once. 
We also restructure the network, which discards the two-decoder structure as well as the multi-task learning module in \cite{zhang2022complex}, in order to make the system parameters more compact. 
In addition, a cross-domain loss function is designed for better generalization in the different conditions. 

The remainder of this paper is organized as follows. We formulate the SAEC problem in Section~\ref{Problem Formulation}. In Section~\ref{Proposed Network}, we introduce the neural network of the proposed framework. The experimental setups are described in Section~\ref{Experiments setups}. We demonstrate and discuss the results of the proposed method in Section~\ref{Results and discussions}. Section~\ref{conclusion} concludes the paper.

\section{Problem formulation}
\label{Problem Formulation}
\subsection{Signal model}

Without loss of generality, we take the common teleconferencing system which has dual-microphone and dual-loudspeaker as example. In the near-end room, the far-end signals $x_{i}(k)$ are generated by a sound source convoluted with room impulse responses (RIRs) $g_{i}(k)$, where $k$ indexes a time sample, $i=1, 2$. The $x_{i}^{nl}(k)$ is a nonlinear distortion of the far-end signals $x_{i}(k)$ that are played by two loudspeakers. The nonlinearity of the power amplifier/loudspeaker is modeled by a 3rd-order polynomial function \cite{zhang2022complex} as follows 
\begin{equation}
x^{nl}\left ( k \right )= 2ax(k)+ax^{2}(k)+x^{3}(k)
\label{eq1}
\end{equation} where $a=\log \left ( \varepsilon/10 \right )+0.1$, and $\varepsilon \in \left [ 2,5 \right ]$.

The echo signals $e_{ij}(k)$ are generated by $x_{i}^{nl}(k)$ convolving with $h_{ij}(k)$ which denotes the echo path from loudspeaker $i$ to microphone $j$, and $j=1, 2$.
\begin{equation}
e_{ij}(k) = x_{i}^{nl}(k)*h_{ij}(k).
  \label{eq2}
\end{equation}where $\ast$ denotes convolution operation.
Considering the effect of reverberation of the near-end room as well, the near-end speech can be expressed by
\begin{equation}
s_j(k) = p(k) * g_{j}(k) = s_{j}^{early}(k) + s_{j}^{late}(k).
  \label{eq3}
\end{equation}
where $p(k)$ is near-end speaker, $s_{j}^{early}(k)$ denotes the early reflections of near-end signal, $s_{j}^{late}(k)$ is the late reverberation reflections. 

Finally, the signal picked up by microphone $j$ is mixed of echo signals $e_{ij}(k)$ and near-end speech $s_{j}(k)$ as follows
\begin{equation}
y_j(k) = \begin{matrix}{\sum_{i=1}^{2}e_{ij}(k)} + s_{j}^{early}(k) + s_{j}^{late}(k)\end{matrix}.
  \label{eq4}
\end{equation}
We assume that the first microphone is always considered as the reference microphone, i.\,e., $j = 1$.
The goal of SAEC is to recover the early near-end component $s^{early}(k)$ from $y(k)$, and other issues like background noise are not considered in this work.

\begin{figure}[t]
\subfigure[DCSM AEC]{   
\begin{minipage}{0.225\textwidth}
\centering    
\includegraphics[scale=0.36]{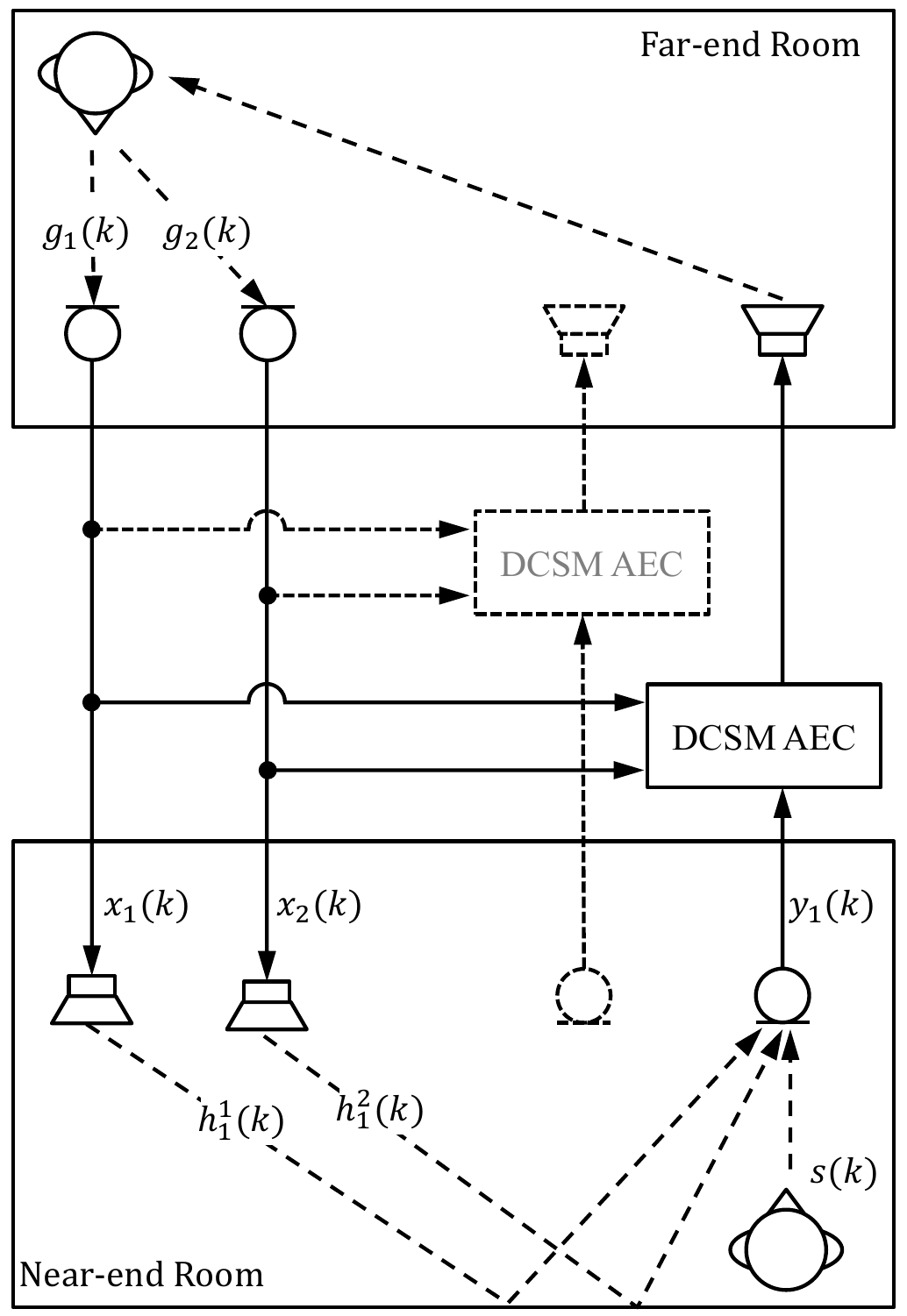}  
\end{minipage}}
\subfigure[DCDM AEC]{ 
\begin{minipage}{0.225\textwidth}
\centering    
\includegraphics[scale=0.36]{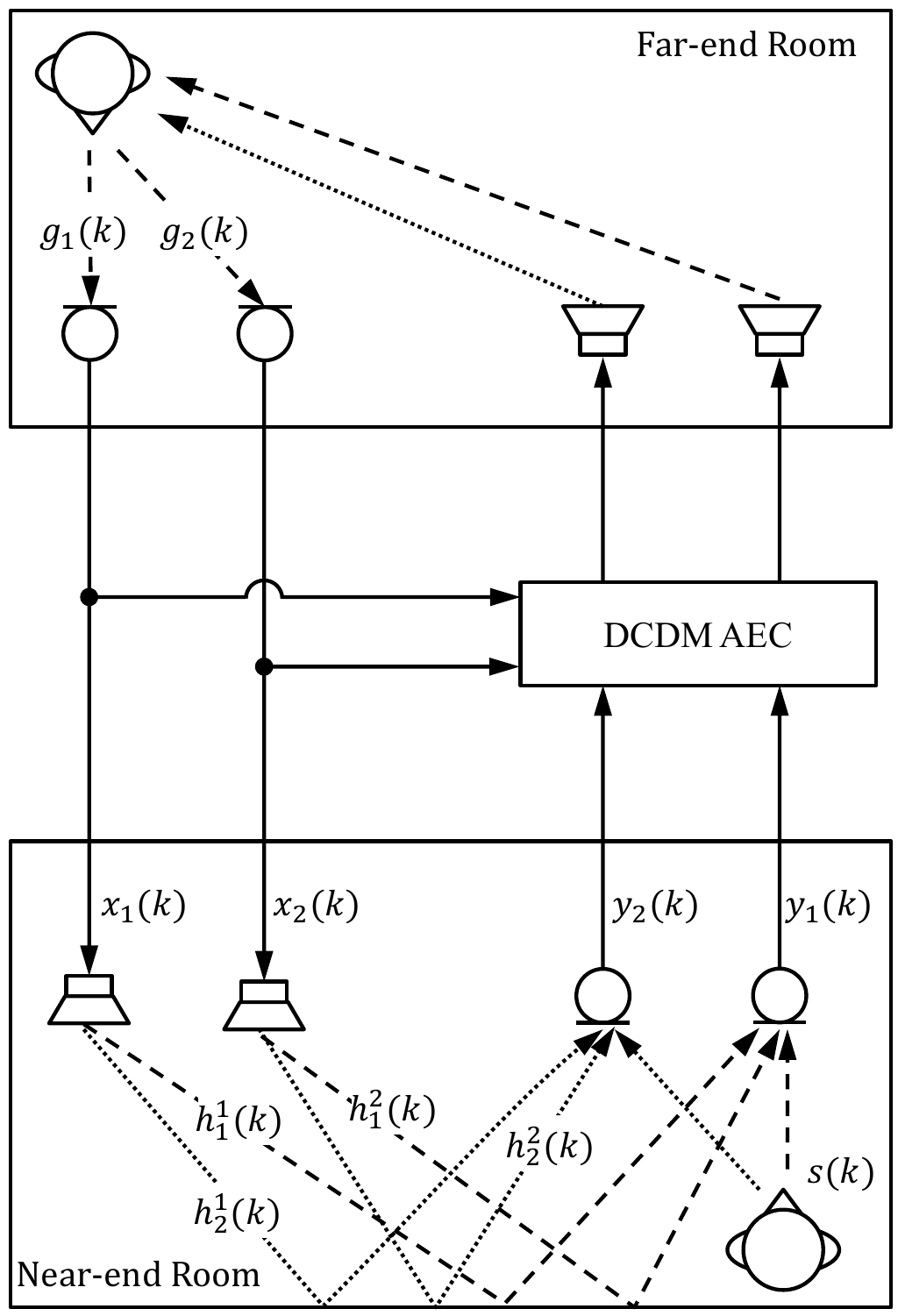}
\end{minipage}}
\caption{ Diagrams of (a) Dual-Channel Single-Microphone AEC setup and (b)
Dual-Channel Dual-Microphone AEC setup.}    
\label{fig:MMCC}    
\end{figure}

\subsection{Dual-Channel Single-Microphone (DCSM) AEC} 
Like traditional scheme, we firstly describe the SAEC problem for two loudspeakers and one microphone setting, because the other microphone needs to be handled in the same way. As illustrated in Fig.~\ref{fig:MMCC}(a), we apply the shared DCSM to each microphone for complex spectral mapping. Specially, during the training, we stack the real and imaginary spectra of one of microphone signal $(Y^j_R, Y^j_I)$ and all far-end signals $(X^i_R, X^i_I)$ as the DCSM inputs (we omit the frame and frequency index for brevity). The complex spectra $S^j_R$ and $S^j_I$ of the early near-end signal $s^{early}_j$ are the learning target, and the model can be expressed as follows
\begin{equation}
\begin{split}
   (S_R^{j}, S_I^{j}) &= DCSM(\Upsilon|X_R^{i}, X_I^{i}, {Y}_{R}^{j}, Y_I^{j}), 
   \\ & i = 1,2.\quad j=1 \text{ or } 2. 
\end{split}
  \label{eq5}
\end{equation}
where $\Upsilon$ is the trainable parameter.
For inference stage, the inverse STFT (iSTFT) is used to synthesize the estimated waveform of the near-end signal $\hat{s}_j(k)$.

\subsection{Dual-Channel Dual-Microphone (DCDM) AEC}
Considering the weakness of the DCSM needs to run two times results in high computational costs. Consequently, one solution would be designed to predict all the near-end signals simultaneously. As shown in Fig.~\ref{fig:MMCC}(b), we proposed DCDM which consists of two loudspeakers and two microphones for joint echo cancellation. Specially, we stack all the real and imaginary spectra of microphone signals $(Y^j_R, Y^j_I)$ and far-end signals $(X^i_R, X^i_I)$ as the DCDM inputs, $S^j_R$ and $S^j_I$ of the near-end signals $s^{early}_j$ as the training target. The model can be expressed by
\begin{equation}
 (S_R^{j}, S_I^{j}) = DCDM(\Psi|X_R^{i}, X_I^{i}, {Y}_{R}^{j}, Y_I^{j}), i,j = 1, 2.
  \label{eq6}
\end{equation}
where $\Psi$ is the trainable parameter.
To the best of our knowledge, this is the first deep learning-based SAEC scheme that can be known as the multi-input multi-output (MIMO) system of SAEC. Since we can obtain an estimate of SAEC only once, the amount of computation is therefore dramatically reduced. 

\begin{figure}[t]
  \centering
  \includegraphics[width=\linewidth]{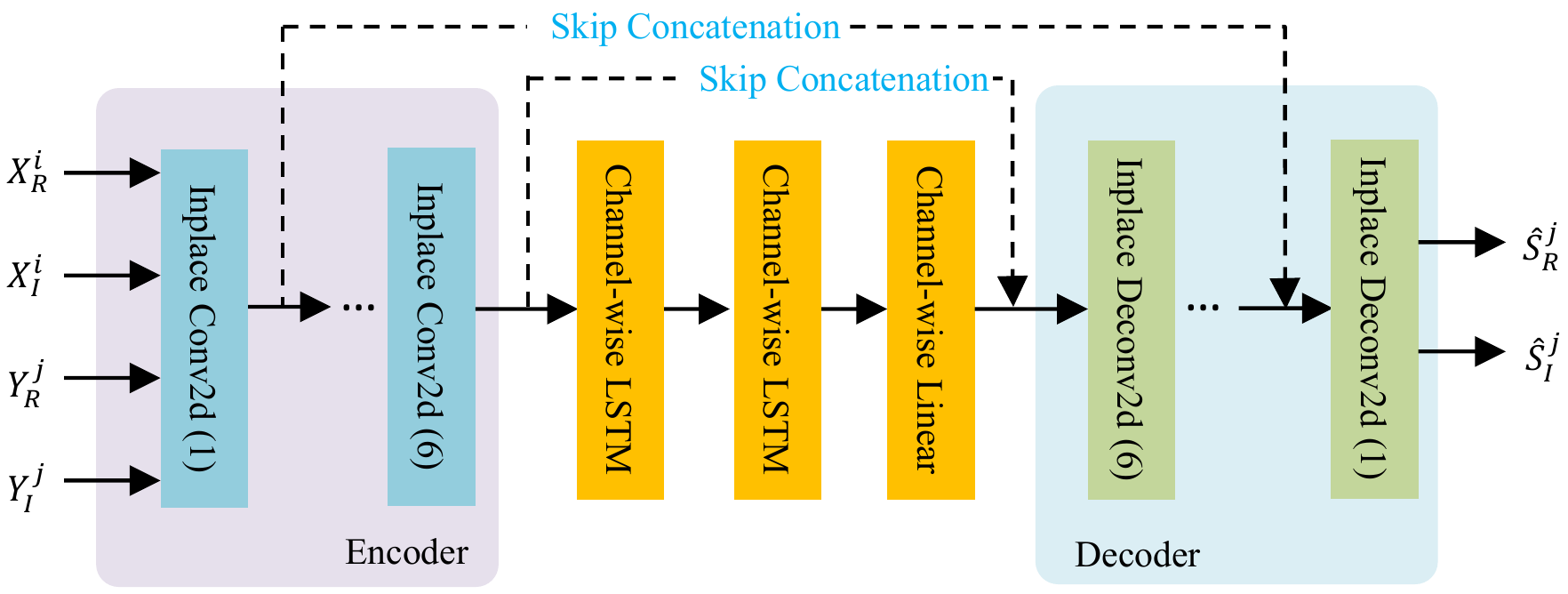}
  \caption{Network architecture of the proposed LCSM. Inplace-Conv2d and Inplace-Deconv2d layer represent the inplace convolution and deconvolution, respectively. The numbers in parenthesis indicate that encoder and decoder are symmetric structures. $(\cdot)_R$ and $(\cdot)_I$ are the real and imaginary spectra, respectively.}
  \label{fig:NET}
\end{figure}

\section{Network structure}
\label{Proposed Network}
\subsection{Inplace convolution and channel-wise LSTM}
The classic CRN \cite{cheng2021deep,zhang2021deep} downsample and upsample operations shrink and extend the features along the frequency dimension in the encoder and decoder, respectively. However, for multi-channel signal processing tasks, the downsampling could confuse spatial cues making LSTM hard to extract the spatial information, since the frequency information tangle with the channel dimension. 

Recently, we proposed inplace convolutional recurrent neural networks, which have been used for speech enhancement and signal-channel AEC tasks \cite{zhang2022complex,liu2021inplace}, and achieved remarkable performance. In this paper, we use a similar network structure for stereo AEC as shown in Fig.~\ref{fig:NET}. Specifically, the inplace operation means that the stride of each convolution/deconvolution layer of the network is set to one, do this for keeping the number of frequency bins unchanged along the frequency dimension \cite{liu2021inplace}. The size of each inplace convolution/deconvolution layer is specified in the $[B, C, T, F]$ format which is short for $[Batchsize, Channels, Time, Frequencies]$. Two Channel-wise LSTM and one Linear layers are shared to extract the spatial information for each bin. The hyperparameters setting are shown in Table~\ref{tb_net} which are given in the ($kernelsize$, $strides$, $outchannels$) format for convolution, and $(hidden\_layer\_size)$ for channel-wise-LSTM and Linear layers. 
A further description of the inplace convolution and channel-wise LSTM is demonstrated in \cite{zhang2022complex,liu2021inplace}. Note that we did not use the two-decoder structure and multi-task learning module employed in \cite{zhang2022complex}, in order to reduce the trainable parameters. 

\begin{table}[t]
	\caption{Descriptions of the hyperparameter settings of DCSM AEC. Note that for DCDM AEC, the input and output channels are set to 8 and 4, respectively.}
	\label{tb_net}
	\footnotesize	
	\renewcommand\tabcolsep{1.8pt}
	\renewcommand\arraystretch{1.3}
\begin{tabular}{|c|c|c|c|}
	\hline
	Layer name                 & Input size & HyperParam. & Output size \\ \hline
	Inplace   Conv2d(1)        & [B, 6, T, F]  & 1$\times$5, (1, 1), 64    & [B, 64, T, F]  \\ \hline
	Inplace   Conv2d(2-6)      & [B, 64, T, F] & 1$\times$5, (1, 1), 64    & [B, 64, T, F]  \\ \hline
	Reshape                    & [B, 64, T, F] & -               & [B$\times$F, T, 64]  \\ \hline
	\begin{tabular}[c]{@{}c@{}}Channel-wise  \\ LSTM$\times$2\end{tabular} & [B$\times$F, T, 64] & 128              & [B$\times$F, T, 128]  \\ \hline
	Channel-wise   Linear                     & [B$\times$F, T, 128] & 64              & [B$\times$F, T, 64]  \\ \hline
	Reshape                    & [B$\times$F, T, 64] & -               & [B, 64, T, F]  \\ \hline
	Inplace   Deconv2d(6-2)  & [B, 128, T, F] & 1$\times$5, (1, 1), 64    & [B, 64, T, F]  \\ \hline
	Inplace   Deconv2d(1)   & [B, 128, T, F] & 1$\times$5, (1, 1), 64    & [B, 2, T, F]   \\ \hline
\end{tabular}
\end{table}

\subsection{Cross-domain loss function}
For complex spectral mapping, the mean-squared error (MSE) loss function may be affected by the energy level of the training samples \cite{gu21_interspeech}. Recent studies \cite{Fusw8168119,wangzhongqiu9018157} demonstrate that combining a magnitude domain mean absolute error (MAE) loss with complex spectrum domain MAE loss is robust, because it better characterizes the distribution of normal data, and the following improved MAE (IMAE) loss function is used.
\begin{equation}
\mathcal{L}_{IMAE}\! =\! \| |\hat{S}| - |S| \|_{1} \!+\! \| \hat{S}_R - {S}_R\|_{1} \!+\! \| \hat{S}_I- {S}_I\|_{1}
\label{eq-mae}
\end{equation}where $\|\cdot\|_{1}$ is the $L_1$ norm, $|\hat{S}|$and $|S|$ denote the magnitude of the estimated signal $\hat{s}(k)$ and target signal $s(k)$, respectively.

To further improve the perceptual quality of the estimated speech, we also adopted the signal-to-distortion ratio (SDR) loss which implicitly integrates phase information in time domain. 
\begin{equation}
\mathcal L_{SDR} = 10\log_{10} \frac{\|s(k)\|^2_2} {\|s(k)-\hat{s}(k)\|^2_2}
\label{eq-sdr}
\end{equation}
where $\hat{s}(k)$ denotes the estimated signal, and $\|\cdot\|_{2}$ is the Euclidean ($L_2$) norm.

Finally, the hybrid multi-domain loss function is obtained: 
\begin{equation}
\mathcal L_{CD} = \mathcal{L}_{IMAE} + \lambda \mathcal L_{SDR}
\label{eq-loss}
\end{equation}
where the parameter $\lambda $ is the weight factor and set to 0.1 according to our experiments. We refer to $\mathcal L_{CD}$ as the cross-domain loss function, as it operates in the complex, magnitude and time-domain.

\vspace{-0.2cm}
\section{Experiments setups}
\label{Experiments setups}
\subsection{Dataset}
We use the raw speech signals from the TIMIT corpus \cite{lamel_speech_1989} for simulating the far-end and near-end speakers as detailed description in \cite{zhang2021deep,zhang2022complex,Zhang2020ARA}. 
Specifically, 20 different simulation rooms of size length, width and height (denote as $l \times w \times h$) $m^{3}$ are designed, where $l$ = [4, 6, 8, 10], $w$ = [5, 7, 9, 11, 13], $h$ = 3. For generating the stereo echo transmission paths, we placed two microphones in the room at positions $(l/2, w/2 + 0.05, h/2)$ m and $(l/2, w/2 - 0.05, h/2)$ m, respectively. 
One near-end speaker and the two loudspeakers are placed at 20 random locations in each room. The distance from the near-end speaker and loudspeakers to the center of the microphones are denoted as $D_{nm}$ and $D_{lm}$, where $D_{nm}$ randomly selected from \{1, 1.2, 1.4\} m and $D_{lm}$ chosen from \{0.5, 0.7, 0.9\} m, respectively. 
It should be noted that the two loudspeakers we used are a vertically symmetrical structure about the center of the microphones. 
The room reverberation time ($T_{60}$) is randomly selected from \{0.2, 0.3, 0.4, 0.5, 0.6\} s, hence, the length of each RIR is $T_{60}\times f_s$, where $f_s$ is signal sampling rate. 
In total, two sets of 6000 RIRs are created using the Image method \cite{allen_image_1979}. In our experiments, we assume that the far-end room and near-end rooms are identical, so the RIRs for the speaker of the both near-end and far-end are shared.
The microphone signal is generated by mixing a near-end signal and two echoes according to a signal-to-echo ratio (SER). In the training stage, SER$\in$[-9, 9] dB and is measured during the double-talk period. We create 30000 pairs of utterances as the training set and 2000 pairs of utterances as the validation set, respectively. 
For the testing experiment, we simulated the untrained room of size $5 \times 6 \times 3$ $m^{3}$, $D_{nm}$, $D_{lm}$ and $T_{60}$ are set to 0.6 m, 1.3 m and 0.35 s, respectively. 
Finally, we generate a test set of 100 pairs of utterances that is entirely distinct from training and validation sets.

\begin{table*}[t]
	\renewcommand\tabcolsep{2.8 pt}
	\renewcommand\arraystretch{1.1}
	\caption{Exhibition of the results under the real office RIR, untrained hard-nonlinearity and music echo conditions with different SERs.}
	\label{tbtot}
	\footnotesize	
	\centering
\begin{tabular}{|c|c|cccccc|cccccc|cccccc|}
\hline
\multirow{4}{*}{} &
   &
  \multicolumn{6}{c|}{Office   RIR} &
  \multicolumn{6}{c|}{Hard-nonlinearity} &
  \multicolumn{6}{c|}{Music   echo} \\ \cline{2-20} 
 &
  Metrics &
  \multicolumn{3}{c|}{ERLE ($\uparrow$)} &
  \multicolumn{3}{c|}{PESQ ($\uparrow$)} &
  \multicolumn{3}{c|}{ERLE ($\uparrow$)} &
  \multicolumn{3}{c|}{PESQ ($\uparrow$)} &
  \multicolumn{3}{c|}{ERLE ($\uparrow$)} &
  \multicolumn{3}{c|}{PESQ ($\uparrow$)} \\ \cline{2-20} 
 &
  SER (dB) &
  -5 &
  0 &
  \multicolumn{1}{c|}{5} &
  -5 &
  0 &
  5 &
  -5 &
  0 &
  \multicolumn{1}{c|}{5} &
  -5 &
  0 &
  5 &
  -5 &
  0 &
  \multicolumn{1}{c|}{5} &
  -5 &
  0 &
  5 \\ \cline{2-20} 
 &
  Mixture &
  — &
  — &
  \multicolumn{1}{c|}{—} &
  1.57 &
  2.00 &
  2.36 &
  — &
  — &
  \multicolumn{1}{c|}{—} &
  1.64 &
  2.06 &
  2.41 &
  — &
  — &
  \multicolumn{1}{c|}{—} &
  1.58 &
  2.05 &
  2.42 \\ \hline
\multirow{5}{*}{DCSM} &
  Yang\cite{yang2012stereophonic} &
  23.43 &
  23.28 &
  \multicolumn{1}{c|}{23.10} &
  2.38 &
  2.75 &
  3.09 &
  14.29 &
  14.25 &
  \multicolumn{1}{c|}{14.14} &
  2.55 &
  2.90 &
  3.21 &
  19.50 &
  19.46 &
  \multicolumn{1}{c|}{19.38} &
  2.74 &
  2.92 &
  3.16 \\
 &
  Cheng \textit{et al.}\cite{cheng2021deep} &
  52.37 &
  52.24 &
  \multicolumn{1}{c|}{51.12} &
  2.41 &
  2.75 &
  3.07 &
  47.96 &
  47.71 &
  \multicolumn{1}{c|}{47.40} &
  2.42 &
  2.76 &
  3.08 &
  \textbf{46.01} &
  \textbf{43.50} &
  \multicolumn{1}{c|}{38.47} &
  2.36 &
  2.68 &
  2.97 \\
 &
  Zhang\cite{zhang2021deep} &
  42.20 &
  41.56 &
  \multicolumn{1}{c|}{39.78} &
  2.54 &
  2.93 &
  3.28 &
  41.44 &
  40.74 &
  \multicolumn{1}{c|}{39.06} &
  2.51 &
  2.91 &
  3.26 &
  33.58 &
  30.59 &
  \multicolumn{1}{c|}{26.60} &
  2.44 &
  2.74 &
  2.95 \\
 &
  LCSM-$\mathcal{L}_{CD}$ &
  44.15 &
  44.29 &
  \multicolumn{1}{c|}{43.40} &
  2.82 &
  3.22 &
  3.57 &
  45.56 &
  45.41 &
  \multicolumn{1}{c|}{44.20} &
  2.89 &
  3.28 &
  3.61 &
  37.27 &
  35.81 &
  \multicolumn{1}{c|}{32.81} &
  2.76 &
  3.04 &
  3.47 \\
 &
  LCSM &
  \textbf{56.24} &
  \textbf{56.27} &
  \multicolumn{1}{c|}{\textbf{55.73}} &
  \textbf{2.84} &
  \textbf{3.25} &
  \textbf{3.59} &
  \textbf{58.58} &
  \textbf{58.32} &
  \multicolumn{1}{c|}{\textbf{57.31}} &
  \textbf{2.91} &
  \textbf{3.31} &
  \textbf{3.63} &
  41.71 &
  40.78 &
  \multicolumn{1}{c|}{\textbf{38.91}} &
  \textbf{2.88} &
  \textbf{3.22} &
  \textbf{3.48} \\ \hline
\multirow{4}{*}{DCDM} &
  Cheng \textit{et al.}\cite{cheng2021deep} &
  50.60 &
  50.66 &
  \multicolumn{1}{c|}{50.45} &
  2.43 &
  2.77 &
  3.09 &
  47.99 &
  47.85 &
  \multicolumn{1}{c|}{47.40} &
  2.45 &
  2.79 &
  3.10 &
  47.89 &
  44.98 &
  \multicolumn{1}{c|}{41.21} &
  2.37 &
  2.71 &
  3.02 \\
 &
  Zhang\cite{zhang2021deep} &
  39.52 &
  39.25 &
  \multicolumn{1}{c|}{38.43} &
  2.45 &
  2.82 &
  3.16 &
  39.88 &
  39.53 &
  \multicolumn{1}{c|}{38.52} &
  2.57 &
  2.92 &
  3.24 &
  33.78 &
  32.49 &
  \multicolumn{1}{c|}{30.10} &
  2.42 &
  2.73 &
  2.99 \\
 &
  LCSM-$\mathcal{L}_{CD}$ &
  47.19 &
  46.82 &
  \multicolumn{1}{c|}{45.66} &
  3.00 &
  3.34 &
  3.63 &
  47.66 &
  47.28 &
  \multicolumn{1}{c|}{46.06} &
  3.04 &
  3.37 &
  3.65 &
  40.01 &
  37.48 &
  \multicolumn{1}{c|}{33.82} &
  3.01 &
  3.28 &
  3.49 \\
 &
  LCSM &
  \textbf{63.78} &
  \textbf{63.12} &
  \multicolumn{1}{c|}{\textbf{61.68}} &
  \textbf{3.02} &
  \textbf{3.35} &
  \textbf{3.64} &
  \textbf{64.02} &
  \textbf{63.50} &
  \multicolumn{1}{c|}{\textbf{62.12}} &
  \textbf{3.05} &
  \textbf{3.38} &
  \textbf{3.66} &
  \textbf{57.46} &
  \textbf{53.43} &
  \multicolumn{1}{c|}{\textbf{48.23}} &
  \textbf{3.06} &
  \textbf{3.32} &
  \textbf{3.52}\\ \hline
\end{tabular}
\end{table*}

\subsection{Baselines and training setups}
We utilize three algorithms as baselines for comparison experiments, there is Yang \cite{yang2012stereophonic} which is a conventional method for addressing the stereo echo suppression problem, and Cheng \textit{et al.} \cite{cheng2021deep} and Zhang \cite{zhang2021deep} are the recent CRN-based methods. The parameter settings are consistent with the description in the original papers.

In this study, all input signals are sampled to 16 kHz, and then divided into frames with 20 ms window length and 50\% overlap, and Hamming window is used. 
The proposed LCSM is trained for 100 epochs by the Adam optimizer \cite{kingma_adam:_2014}, and the batch size is set to 4. The initial learning rate is 1e-3, and dropped by a factor of 0.3 when validation loss does not improve for three epochs. 

\section{Results and discussions}
\label{Results and discussions}
Two general metrics are employed to measure the experimental results, there is the perceptual evaluation of speech quality (PESQ) \cite{itu-t_862perceptual_2001} to measure the near-end speech quality during the double-talk period, and the echo return loss enhancement (ERLE) \cite{enzner_acoustic_2014} to evaluate the echo reduction during the single-talk period, respectively. For the convenience of comparing the performance of DCSM AEC and DCDM AEC, our results are all the estimated near-end signals of the reference microphone.

First, we evaluate the performance of the algorithms on three common unmatched conditions. a) Real office RIR scenario selected from the Aachen Impulse Response database \cite{jeub2009binaural}. b) Unseen nonlinear distortion scenario which utilizes the $hard$-$clipping$ function to simulate the hardware distortions as described in \cite{stenger2000adaptation,malik_state-space_2012}, and the clipping threshold of the input signal is set to 0.7 in this work. 
c) Music echoes scenario, where far-end signals are music randomly chosen from MUSAN database \cite{snyder2015musan}, is also a very common echo in practice.

The average scores of ERLE and PESQ for all methods in different scenarios are demonstrated in Table \ref{tbtot}. Among these methods, LCSM-$\mathcal{L}_{CD}$ represents LCSM which uses MSE loss(used in Zhang) instead of $\mathcal{L}_{CD}$ for ablation studies. 
From the table, we can infer the following conclusions. 
1) The proposed LCSM achieves the best scores from DCSM to DCDM settings in terms of ERLE and PESQ in most conditions, and the complexity is greatly reduced.
2) The LCSM yields better performance than LCSM-$\mathcal{L}_{CD}$, especially in terms of ERLE. It implies that the L-$\mathcal{L}_{CD}$ significantly improves the generalization ability.
3) Because it is a statistical model of the signal, Yang achieves stable performance in terms of PESQ during the double-talk periods but yields lower scores in terms of ERLE which denotes the echo cancellation ability is limited. 
4) Although Cheng \textit{et al.} achieves impressive ERLE scores in all scenarios, the near-end signal quality is distorted seriously since the phase information is ignored.
5) The performance of LCSM-$\mathcal{L}_{CD}$ is greatly improved over Zhang, especially in terms of PESQ. And the results confirm that the inplace convolution operation and the temporal modeling of each frequency bin are valid, as we expected.

Further, we also conduct the condition in which the near-end is music but the far-end is speech. 
Comparisons of all methods are depicted in Table~\ref{near-music}. We utilize SDR as the metric because PESQ is hard to measure the music signal quality. 
\begin{table}[t]
	\renewcommand\tabcolsep{2.3 pt}
	\renewcommand\arraystretch{1.0}
	\caption{Comparison of the proposed LCSM with other algorithms under the far-end is speech while near-end is music signal condition with different SERs.}
	\label{near-music}
	\footnotesize	
	\centering
\begin{tabular}{|c|c|cccccc|}
\hline
                      &               & \multicolumn{6}{c|}{Far-end(speech), Near-end(music)}                                 \\ \cline{2-8} 
                      & Metrics       & \multicolumn{3}{c|}{ERLE ($\uparrow$)}                  & \multicolumn{3}{c|}{SDR ($\uparrow$)} \\ \cline{2-8} 
                      & SER (dB)          & -5    & 0     & \multicolumn{1}{c|}{5}     & -5     & 0      & 5      \\ \hline
\multirow{5}{*}{DCSM} & Yang\cite{yang2012stereophonic}    & 14.45 & 14.37 & \multicolumn{1}{c|}{14.21} & -0.01  & 0.38   & 0.51   \\
                      & Cheng \textit{et al.}\cite{cheng2021deep}        & 49.99 & 47.32 & \multicolumn{1}{c|}{44.15} & 5.04   & 7.52   & 10.46  \\
                      & Zhang\cite{zhang2021deep}         & 41.87 & 40.03 & \multicolumn{1}{c|}{37.13} & 5.53   & 7.75   & 10.06  \\
                    & LCSM-$\mathcal{L}_{CD}$        & 46.25 & 44.66 & \multicolumn{1}{c|}{42.23} & 8.79   & 11.49   & 13.14  \\
                      & LCSM    & \textbf{56.99} & \textbf{54.21} & \multicolumn{1}{c|}{\textbf{51.07}} & \textbf{9.20}   & \textbf{12.06}  & \textbf{14.80}  \\ \hline
\multirow{4}{*}{DCDM} & Cheng \textit{et al.}\cite{cheng2021deep}        & 49.04 & 46.97 & \multicolumn{1}{c|}{43.79} & 5.13   & 7.42   & 10.07  \\
                      & Zhang\cite{zhang2021deep}         & 40.03 & 38.98 & \multicolumn{1}{c|}{37.06} & 5.59   & 7.95   & 10.44  \\
                    & LCSM-$\mathcal{L}_{CD}$         & 46.97	 & 45.41 & \multicolumn{1}{c|}{43.12} & 9.31   & 11.60   & 13.85  \\
                    & LCSM         & \textbf{59.41} & \textbf{56.56} & \multicolumn{1}{c|}{\textbf{53.45}} & \textbf{10.22} & \textbf{12.59} & \textbf{15.04} \\ \hline
\end{tabular}
\end{table}
\begin{figure}[t]
	\centering	
	\subfigure[Microphone signals]{
		\begin{minipage}[]{0.225\textwidth}
			\includegraphics[width=\textwidth, height= 55 pt]{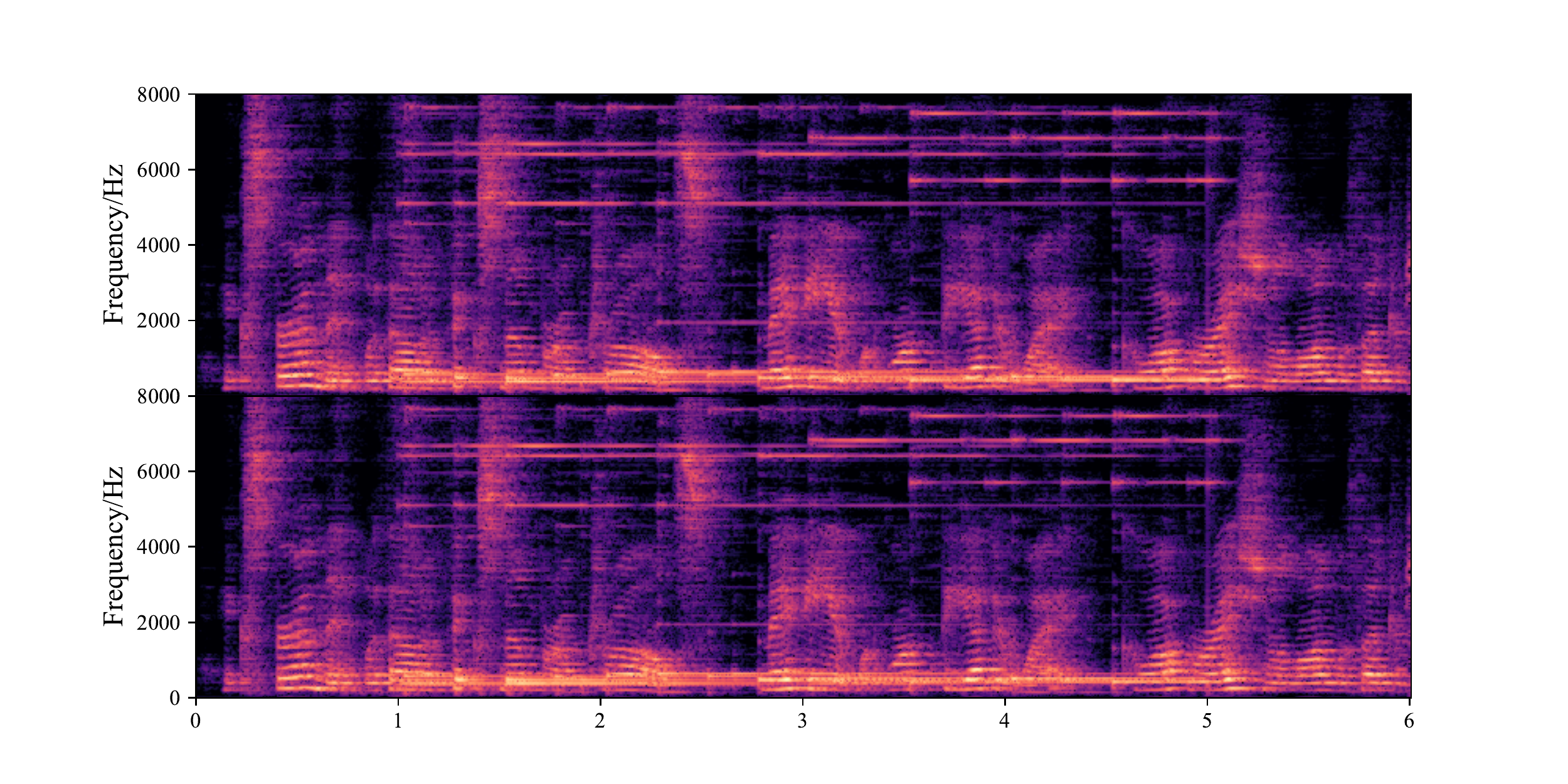}
	\end{minipage}}	
	\subfigure[Cheng \textit{et al.} estimated signals]{
		\begin{minipage}[]{0.225\textwidth}
			\includegraphics[width=\textwidth, height= 55 pt]{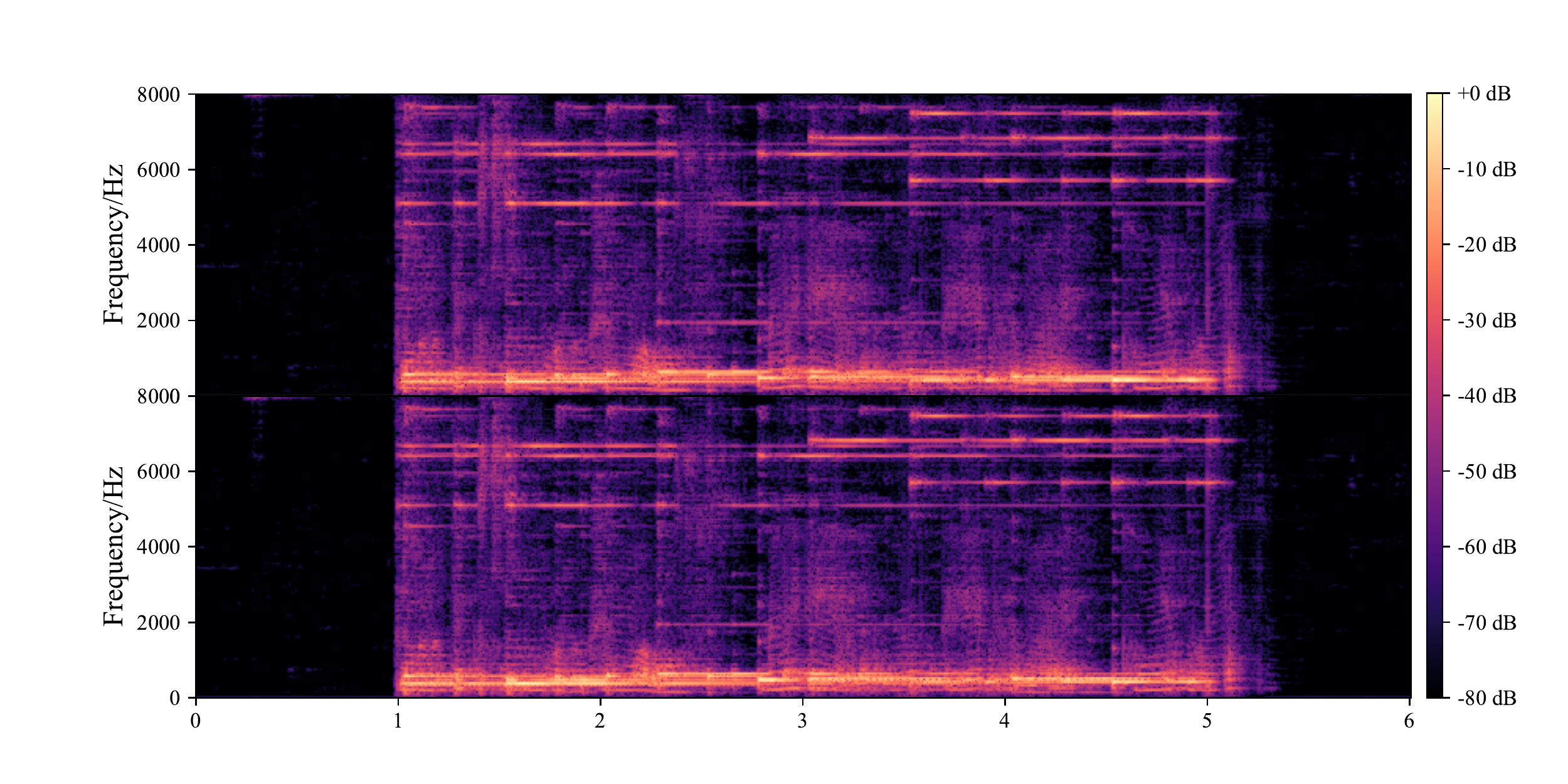}
	\end{minipage}}	
	\subfigure[Zhang estimated signals]{
		\begin{minipage}[]{0.225\textwidth}
			\includegraphics[width=\textwidth, height= 55 pt]{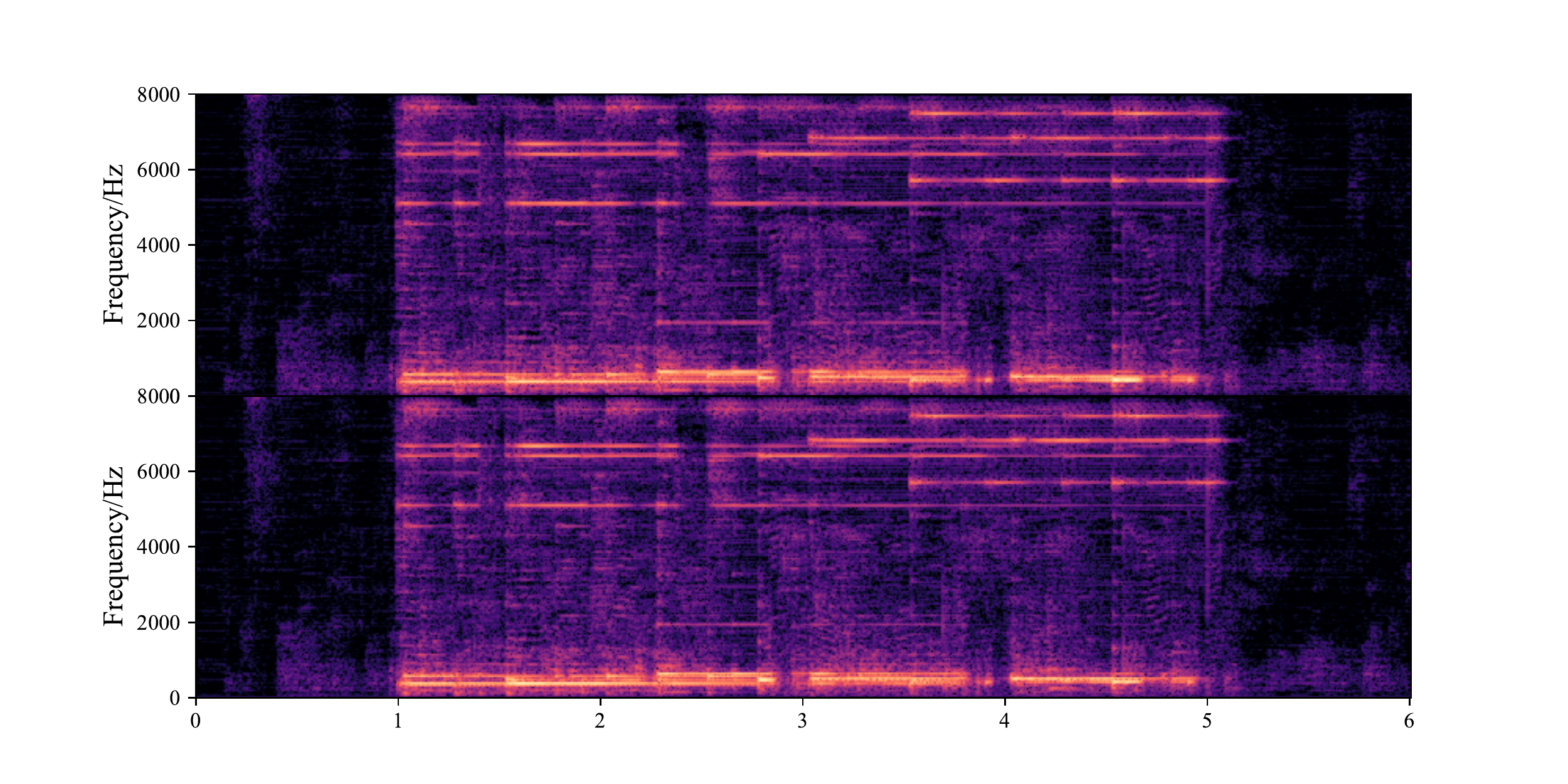}
	\end{minipage}}
	\subfigure[LCSM estimated signals]{
		\begin{minipage}[]{0.225\textwidth}
			\includegraphics[width=\textwidth, height= 55 pt]{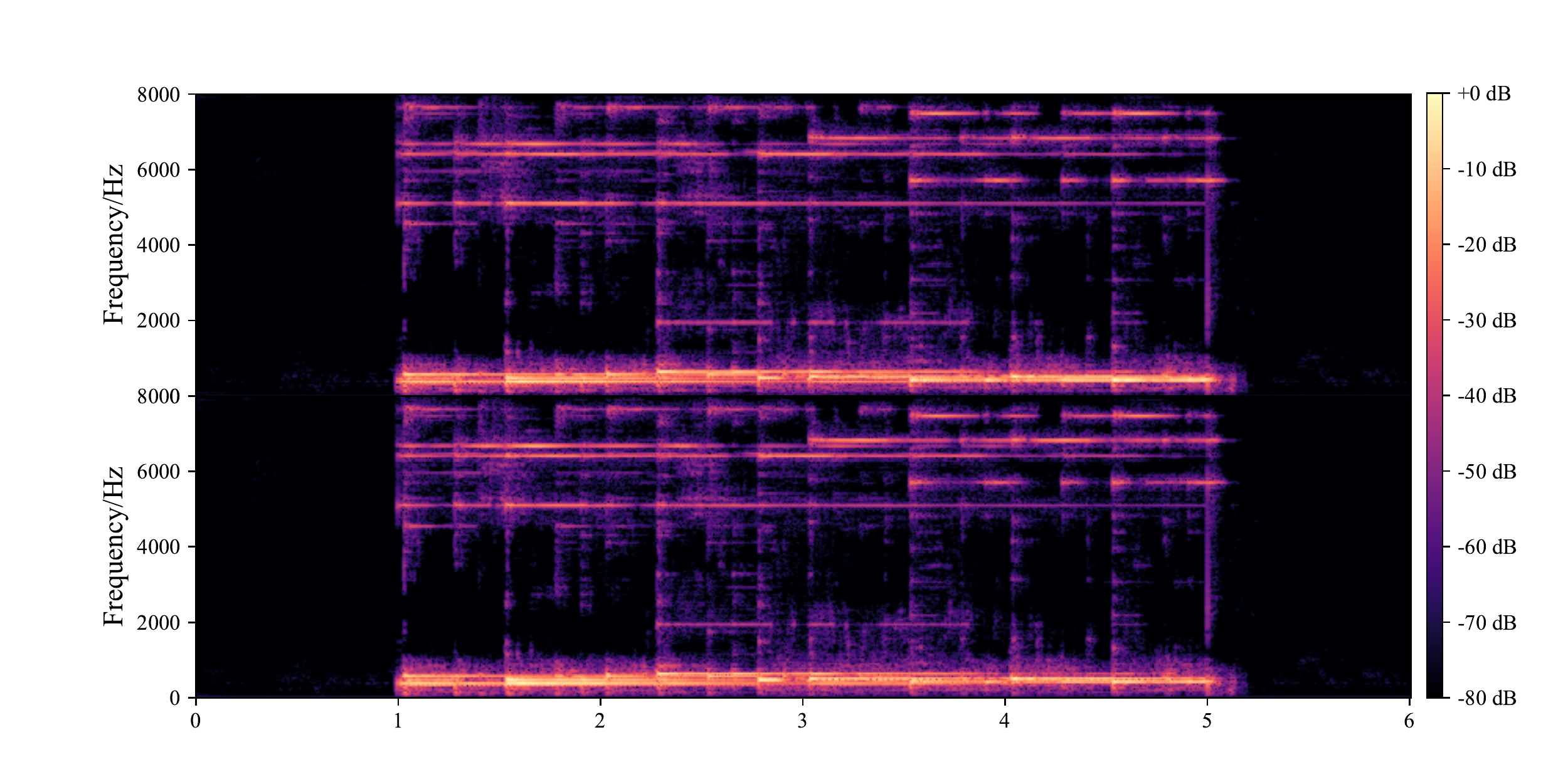}
	\end{minipage}}
	\caption{Presentation of spectra of different DCSM AEC methods at SER = 0 dB. Note that the top of image (a) is the first microphone signal and the bottom is the second microphone signal. }
	\label{spectra}

\end{figure}
\begin{table}[t]
	\renewcommand\tabcolsep{1 cm}
	\renewcommand\arraystretch{1.0}
	\caption{Results of the number of trainable parameters for the deep learning-based SAEC. (unit is Million).}
	\label{params}
	\footnotesize	
	\centering
\begin{tabular}{|c|c|c|}
\hline
Methods & Parameters \\ \hline
Cheng \textit{et al.}\cite{cheng2021deep}  & 12.99 M     \\
Zhang\cite{zhang2021deep}   & 9.07 M      \\
LCSM    & 0.55 M    \\ \hline
\end{tabular}
\end{table}
As shown in Table~\ref{near-music}, the proposed LCSM has significant advantages over other methods in both DCSM and DCDM settings. Take SER = 0 dB in DCSM setting for example, the near-end music signal estimated by the LCSM improves SDR by 4.31 dB and improves ERLE by 14.18 dB over Zhang, respectively. 
Clearer spectra of different signals are exhibited in Fig.~\ref{spectra}. 
Some audio samples of this scenario can be found in this page\footnote[1]{https://chenggangzhang.github.io/LCSM-StereoAEC}.

Considering the limitation of memory resources, we also compare the number of trainable parameters for the deep learning-based approach. Please note that the difference between DCSM and DCDM in terms of parameters is not very obvious. From Table~\ref{params}, we can find that the parameters of the framework are only 0.55 M which is over 16x less than Zhang. Furthermore, the proposed LCSM is a causal system that is essential for real-time applications.

\section{Conclusion}
\label{conclusion}
In this paper, a lightweight end-to-end framework is presented for solving the SAEC problem. We employ inplace convolution and channel-wise temporal modeling to keep the vital information of the near-end signal. Moreover, a cross-domain loss function is adopted for complex spectral mapping. It is shown in the performance evaluation that the proposed method has good generalization ability. In the following studies, we plan to investigate the performance of the proposed framework in real stereophonic scenarios.

\section{Acknowledgements}
\vspace{-0.1cm}
The authors would like to thank Hao Zhang for his helpful discussions in the early stage. This research was supported by the China National Nature Science Foundation (No.61876214).

\bibliographystyle{IEEEtran}
\bibliography{AEC_paper}
\end{document}